\documentclass[preprint2]{aastex6}

\usepackage{amsmath,multirow,rotating}
\usepackage[graphicx]{realboxes}
\hypersetup{breaklinks=true,colorlinks=true,citecolor=blue,linkcolor=blue,urlcolor=blue}

\shorttitle{Star formation, SNe, iron and $\alpha$}
\shortauthors{Maoz \& Graur}

\begin{document}

\title{Star formation, supernovae, iron, and $\alpha$: consistent cosmic and Galactic histories}

\author{Dan Maoz}
\affil{School of Physics and Astronomy, Tel-Aviv University, Tel-Aviv 6997801, Israel; \url{maoz@astro.tau.ac.il}}

\author{Or Graur\altaffilmark{1}}
\affil{Harvard-Smithsonian Center for Astrophysics, 60 Garden St., Cambridge, MA 02138, USA; \url{or.graur@cfa.harvard.edu}}
\affil{Department of Astrophysics, American Museum of Natural History, New York, NY 10024, USA}

\altaffiltext{1}{NSF Astronomy and Astrophysics Postdoctoral Fellow.}

\begin{abstract}
\noindent Recent versions of the observed cosmic star-formation history (SFH) have resolved an inconsistency with the stellar mass density history. We show that the revised SFH also scales up the delay-time distribution (DTD) of Type Ia supernovae (SNe Ia), as determined from the observed volumetric SN Ia rate history, aligning it with other field-galaxy SN Ia DTD measurements. The revised-SFH-based DTD has a $t^{-1.1 \pm 0.1}$ form and a Hubble-time-integrated production efficiency of $N/M_\star=1.3\pm0.1$ SNe Ia per $1000~{\rm M_\odot}$ of formed stellar mass. Using these revised histories and updated empirical iron yields of the various SN types, we re-derive the cosmic iron accumulation history. Core-collapse SNe and SNe Ia have contributed about equally to the total mass of iron in the Universe today. We find the track of the average cosmic gas element in the [$\alpha$/Fe] vs. [Fe/H] abundance-ratio plane. The track is broadly similar to the observed main locus of Galactic stars in this plane, indicating a Milky Way (MW) SFH similar in form to the cosmic one. We easily find a simple MW SFH that makes the track closely match this stellar locus. Galaxy clusters appear to have a higher-normalization DTD. This cluster DTD, combined with a short-burst MW SFH peaked at $z=3$, produces a track that matches remarkably well the observed ``high-$\alpha$'' locus of MW stars, suggesting the halo/thick-disk population has had a galaxy-cluster-like formation mode. Thus, a simple two-component SFH, combined with empirical DTDs and SN iron yields, suffices to closely reproduce the MW's stellar abundance patterns. 

\end{abstract}


\section{Introduction}
\label{sec:intro}

A powerful approach to observational cosmology has been the direct quantitative inventory of star formation rates, accumulated stellar mass, and the rates of supernovae (SNe), all as a function of cosmic
time (see \citealt{2014ARA&A..52..415M}, hereafter MD14, for a recent review). Increasingly sophisticated observations and analyses over the past decades have brought into improved focus the star-formation-rate density as a function of redshift, referred to as the cosmic star formation history (SFH), and the stellar mass density evolution, both to redshifts of $z\sim 8$. Integration over time of the SFH, however, had overpredicted the accumulated stellar density by factor of a few (e.g., \citealt{2006ApJ...651..142H}). This led to suggestions (e.g., \citealt{2008MNRAS.385..687W}) that early star formation proceeded with a ``top heavy'' initial mass function (IMF), so that relatively few low-mass stars remained after the massive stars, whose measurements drive the star formation rate (SFR) estimates, completed their stellar evolution.
 
A possibly related problem emerged from the measurement of core-collapse (CC) SN rates versus redshift, out to $z\sim 1$. Since all stars with initial masses $\gtrsim 8~{\rm M_\odot}$ are thought to explode within tens of Myr after star formation, i.e., essentially instantaneously on cosmic timescales, the volumetric CC SN rate (i.e., number of SNe per unit time per unit cosmological volume) is expected to follow the SFH almost exactly, with a scaling determined by the number of high-mass stars formed per unit total stellar mass formed. For example, for a \citet{2001MNRAS.322..231K} IMF (which we will assume consistently throughout this paper; a broken power-law distribution in mass, with exponents of $-1.3$ between 0.1 and 0.5~M$_\odot$, and $-2.3$ above 0.5~M$_\odot$), the scaling is $0.010 \pm 0.002~{\rm M_\odot}^{-1}$, i.e., about one CC SN for every $100~{\rm M_\odot}$ of stars formed (with the error coming from the uncertainty in the minimal progenitor mass of a CC SN, between 7.5~M$_\odot$ and 10~M$_\odot$; \citealt{Smartt2009review}). However, this scaling of the SFH overpredicted the observed CC SN rates, and led to suggestions (e.g., \citealt{2011ApJ...738..154H}) that perhaps a significant fraction of massive stars do not explode as SNe, but rather collapse and become black holes without explosions. A more plausible solution that has emerged is that SN surveys systematically miss CC SNe in star-forming, dusty galaxies. CC SN surveys that attempt to correct for the fraction of missing CC SNe generally find good agreement between the corrected rates and the SFH (see MD14 and, most recently, \citealt{2015MNRAS.450..905G} and \citealt{2015ApJ...813...93S}). An infrared survey attempting to measure the fraction of missing CC SNe directly is ongoing (e.g., \citealt{2008ApJ...689L..97K,2012ApJ...744L..19K}).

As summarised in MD14, a more mundane solution also to the problem of the predicted stellar density excess comes from the realization that most works earlier than about 2007 overestimated the SFR over the redshift range $z\sim 2$--$7$. The main causes of the SFR overestimates were over-corrections of ultraviolet-based SFRs for dust absorption (e.g., a factor--$4.5$ correction at $z=2$ in \citealt{2006ApJ...647..128E}, which went down to a factor $\sim 2$ in \citealt{2009ApJ...692..778R}), and over-corrections for the unseen faint end of the galaxy luminosity function. These corrections are now superfluous or less important, with the availability of far-infrared-based SFR estimates that circumvent the need for dust corrections, and deeper observations that probe directly more of the faint end. With the more recent SFHs, the predicted stellar mass density history is in much better agreement with the observations.

In this paper, we show that two additional paradoxes are resolved when a modern SFH is adopted. The delay-time distribution (DTD) of Type Ia SNe (SNe Ia), obtained by comparing the volumetric SN Ia rate history to the SFH, becomes consistent with other DTD measurements (which are not cosmic-SFH-based); and the contribution of SNe~Ia to the recent and present-day cosmic budget of iron becomes equal to that of CC~SNe (as opposed to a sub-dominant SN~Ia role when assuming older versions of the SFH), as deduced also for the SN~Ia contributions to the Sun's iron content, based on the abundance ratios of $\alpha$ elements and iron in Galactic stars. Furthermore, the SN Ia DTD, now reliably determined, when combined with empirical SN iron yields, permits a derivation of the effective SFH of the Milky Way's (MW) various components that reproduces remarkably well the latest precision data on the distribution of stellar abundances.

\section{The SN Ia delay time distribution}
\label{sec:DTD}

The DTD of SNe Ia is the hypothetical distribution of delay times between the formation of a single stellar population of unit total mass and the explosion of some members of the population, or their
descendants, as SNe Ia. The DTD is of fundamental importance. It dictates the rate at which SNe Ia contribute their nucleosynthetic and energy output to individual galaxies and to the Universe as a whole. The form of the DTD, in turn, is set by the progenitor population of SNe Ia, and by the stellar and binary evolution of that population. Different SN Ia progenitor models therefore make different DTD predictions (e.g., \citealt{2013FrPhy...8..116H,2014A&A...563A..83C,2014A&A...562A..14T}), and the measurement of an observational DTD is a useful tool for discriminating among progenitor models and their physics.

\citet[hereafter M14]{2014ARA&A..52..107M} have reviewed a number of recent observational determinations of the SN~Ia DTD, which we briefly recap here. \citet{2008PASJ...60.1327T} compared SN Ia rates in $z=0.4$--$1.2$ early-type galaxies of various deduced ages. Due to the problematics of using a galaxy's luminosity-weighted age as a proxy for the SN Ia progenitor age (see \citealt{Maoz2012review} for details), we do not use these measurements here. \citet{Maoz2010clusters} found the DTD by combining previous measurements of SN Ia rates in samples of massive galaxy clusters at redshifts $0<z<1$, and deriving the SN~Ia rate per formed unit stellar mass as a function of time since the main epoch of star formation in the clusters. \citet{Maoz2010loss} derived the DTD by using the individual galactic SFHs, determined via modeling of the galaxy spectra by \citet{Tojeiro2007}, of the nearby galaxies in the Lick Observatory Supernova Survey (LOSS; \citealt{2011MNRAS.412.1419L}) which were also observed as part of the Sloan Digital Sky Survey (SDSS; \citealt{2000AJ....120.1579Y}), and comparing them to the number of SNe~Ia discovered in each galaxy in the course of the LOSS survey. \citet{Maoz2012sdss} recovered the DTD by applying the same technique to a sample of galaxies in the SDSS and to the SNe Ia that they hosted in the SDSS II SN survey \citep{2008AJ....135..338F}. \citet{GraurMaoz2013} did likewise, but for the full SDSS spectroscopic sample of $\sim700,000$ galaxies in Data Release 7 \citep{SDSS-DR7}, comparing their individual SFHs to the SNe Ia that some of them happened to host just when their spectra were being obtained, and hence the SN~Ia spectrum was included in the fiber aperture, on top of the galaxy spectrum. Finally, \citet{Graur2011} and \citet{Graur2014} derived the DTD by finding the function that, when convolved with the cosmic SFH, best fits the volumetric SN~Ia rate measurements from an assortment of untargeted field SN surveys at $0<z<2$.

As summarized by M14, all of these diverse DTD measurements, based on observations of diverse stellar populations at diverse redshifts, and using different methodologies to recover the DTD, have found a
consistent DTD that depends on delay time roughly as $\sim t^{-1}$. Such a time dependence is expected generically in models where the main parameter setting the time until explosion is the separation of the SN Ia progenitor binary system, and where there is a steep dependence of this time-to-explosion on the said separation. In particular, the traditional ``double-degenerate'' SN~Ia progenitor model \citep{Iben1984,Webbink1984}, in which a close double-white-dwarf binary loses energy to gravitational waves until it merges and explodes, predicts such a $\sim t^{-1}$ form for the DTD.

Apart from their $\sim t^{-1}$ form, most of the observationally-derived DTDs also agree in their normalization, which can be expressed in terms of the Hubble-time-integrated number of SNe~Ia that will eventually explode from a formed unit mass of new stars. \citet{Maoz2010loss} and \citet{Maoz2012sdss} found broadly consistent values of $N/M_\star = (2.3 \pm 0.6)\times10^{-3}$ and $(1.30 \pm 0.15)\times10^{-3}~{\rm M}_\odot^{-1}$, respectively. (\citet{GraurMaoz2013} recovered only the late-time component of the DTD, which can also be used to constrain the DTD normalization.\footnote{A conversion error in \citet{GraurMaoz2013} in the DTD scaling between the \citet{2001MNRAS.322..231K} IMF assumed by \citet{Tojeiro2007} and the \citet{2003ApJS..149..289B} ``diet'' Salpeter IMF used by \citet{Maoz2010clusters} has been corrected here. For an old stellar population  with luminosity dominated by solar-mass stars, the diet-Salpeter IMF has (by definition) 0.7 the mass of the Salpeter IMF. The \citet{2001MNRAS.322..231K} IMF, when matched to the Salpeter IMF at 1~M$_\odot$, has 0.76 the mass of the Salpeter IMF. An old stellar population with a given luminosity thus has a 9\% higher initial mass when assuming a \citet{2001MNRAS.322..231K} IMF compared to a  \citet{2003ApJS..149..289B} ``diet'' Salpeter IMF.} In galaxy clusters, \citet{Maoz2010clusters} found a DTD normalization higher by a factor of $\sim 2$--$4$. This was evidenced by the high levels of the DTD at long delays, based directly on observed cluster SN~Ia rates, but also by the high observed iron-to-stellar mass ratio in clusters, which indicates $N/M_\star\approx (3.3$--$8.1)\times10^{-3}~{\rm M}_\odot^{-1}$. The cause of this is still unclear (see M14 for a discussion), and we return to this issue in the context of the observed stellar abundances in our Galaxy in Section~\ref{sec:iron}. Perhaps more puzzling, however, the DTD normalization found by \citet{Graur2011} and \citet{Graur2014} was a factor of $\sim 2$ \emph{lower} than the field-galaxy-based measurements. This was hard to understand given that the monitored and host galaxies in the targeted surveys were not very different from those in the untargeted surveys, particularly the $z\lesssim 0.7$ hosts and their SNe that mostly drive the DTD results from the volumetric SN rates. We now show that the more recently measured cosmic SFHs resolve this discrepancy.

\begin{figure}
 \centering
 \includegraphics[width=0.47\textwidth]{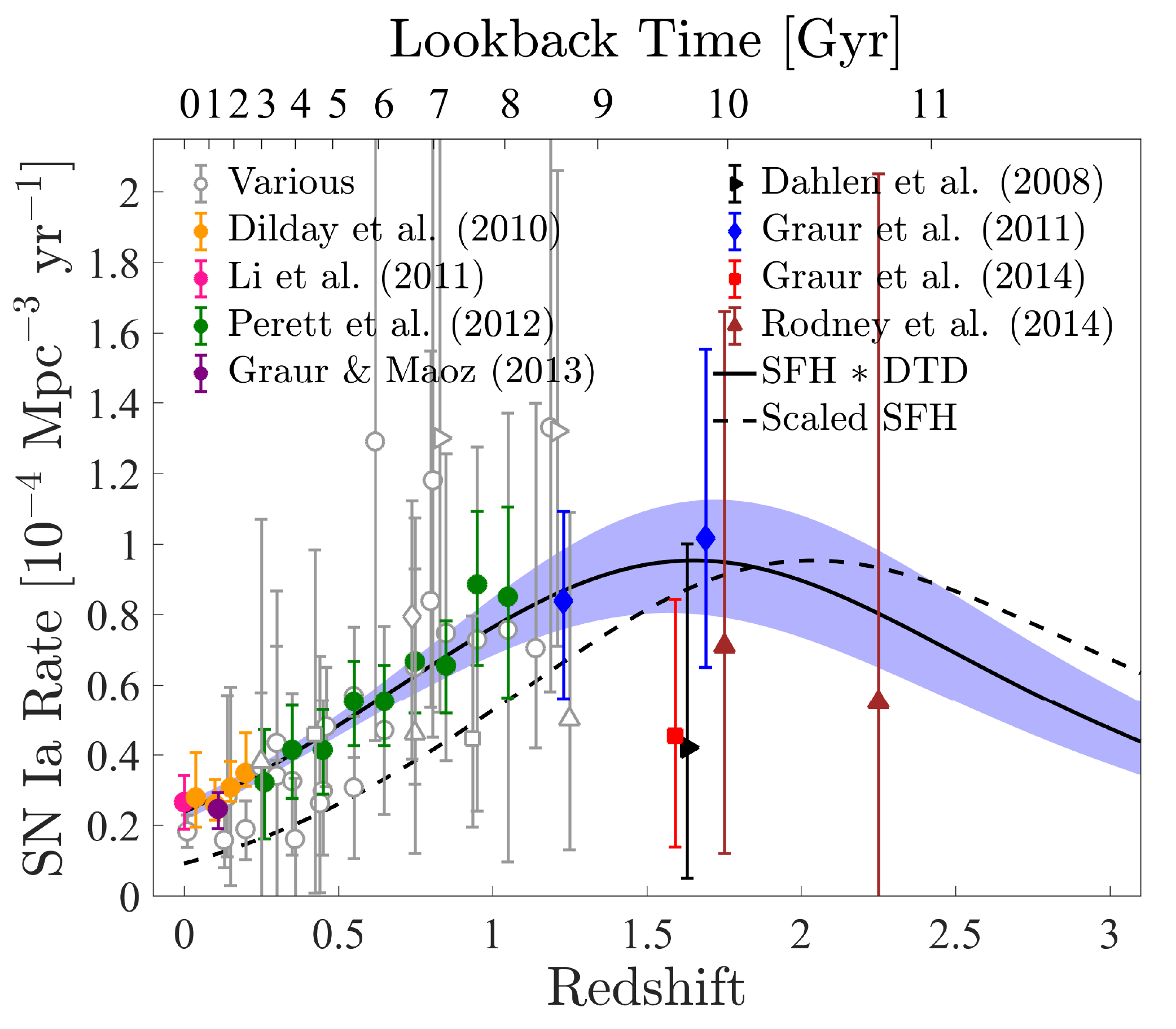}
 \caption{Volumetric field SN~Ia rates as a function of redshift. The solid curve gives the best-fit model SN~Ia rate evolution obtained by convolving the SFH of MF17 with a DTD that is a $t^{\alpha}$ power
   law with $\alpha=-1.1\pm 0.1$, shown in Fig.~\ref{fig:dtd}. To illustrate the relation between the SFH and the SN~Ia rate, we plot (dashed curve) the MF17 SFH, scaled to have its maximum at the same level as that of the best-fitting SN rate evolution. All measurements at $z<1$ are denoted with circles, while measurements at $z\ge1$ are denoted with various symbols, as marked. The most precise measurements at each redshift are highlighted with filled symbols; at $z>1.5$, all measurements are shown (Table~\ref{table:rates}). Measurements with lower precision at $z<1$ are listed as ``various'' and were taken from \citet{cappellaro1999,hardin2000,2002ApJ...577..120P,2003ApJ...594....1T,blanc2004,botticella2008,horesh2008,2010ApJ...723...47R,li2011rates}; and \citet{melinder2012}.}
 \label{fig:rates}
\end{figure}

We represent the SFH with the best-fitting functional representation by \citet[hereafter MF17]{2017ApJ...840...39M}, which updates the SFH of MD14 for recent measurements at $z>4$, and assumes the same
\citet{2001MNRAS.322..231K} IMF that we consistently use: 
\begin{equation}\label{eq:eq1}
 \psi(z)=0.01 \times \frac{(1+z)^{2.6}}{1+[(1+z)/3.2]^{6.2}}~{\rm
 M}_\odot ~{\rm yr}^{-1}~ {\rm Mpc}^{-3}.
\end{equation}
Figure~\ref{fig:rates} shows the compilation of volumetric field SN~Ia rates as a function of redshift, as already shown in \citet{Graur2014}. The solid curve gives the best-fit model SN~Ia rate evolution obtained by convolving the MF17 SFH with a range of trial DTDs. The trial DTDs have zero rates from time $t=0$ to $t=40$~Myr (corresponding to the time of formation of the first white dwarfs in a stellar population). They then rise instantaneously to a maximum level that is a free parameter, and then decline according to a power law in time, with a power-law index, $\alpha$, that is a free parameter. When fitting only the more precise among the measurements in Figure~\ref{fig:rates} (shown with filled symbols and summarized in Table~\ref{table:rates}), the best fit is $\alpha=-1.07^{+0.09}_{-0.08}$ ($\chi^2=7$ for 19 degrees of freedom). Fitting all the measurements shown in Figure~\ref{fig:rates} gives a similar result: $\alpha=-1.10^{+0.08}_{-0.07}$ ($\chi^2=36$ for 52 degrees of freedom). The Hubble-time-integrated SN~Ia production efficiency is $N/M_\star=1.3\pm 0.1$ SNe per $1000~{\rm M_\odot}$ of stellar mass formed.

\begin{figure}
 \centering
 \includegraphics[width=0.47\textwidth]{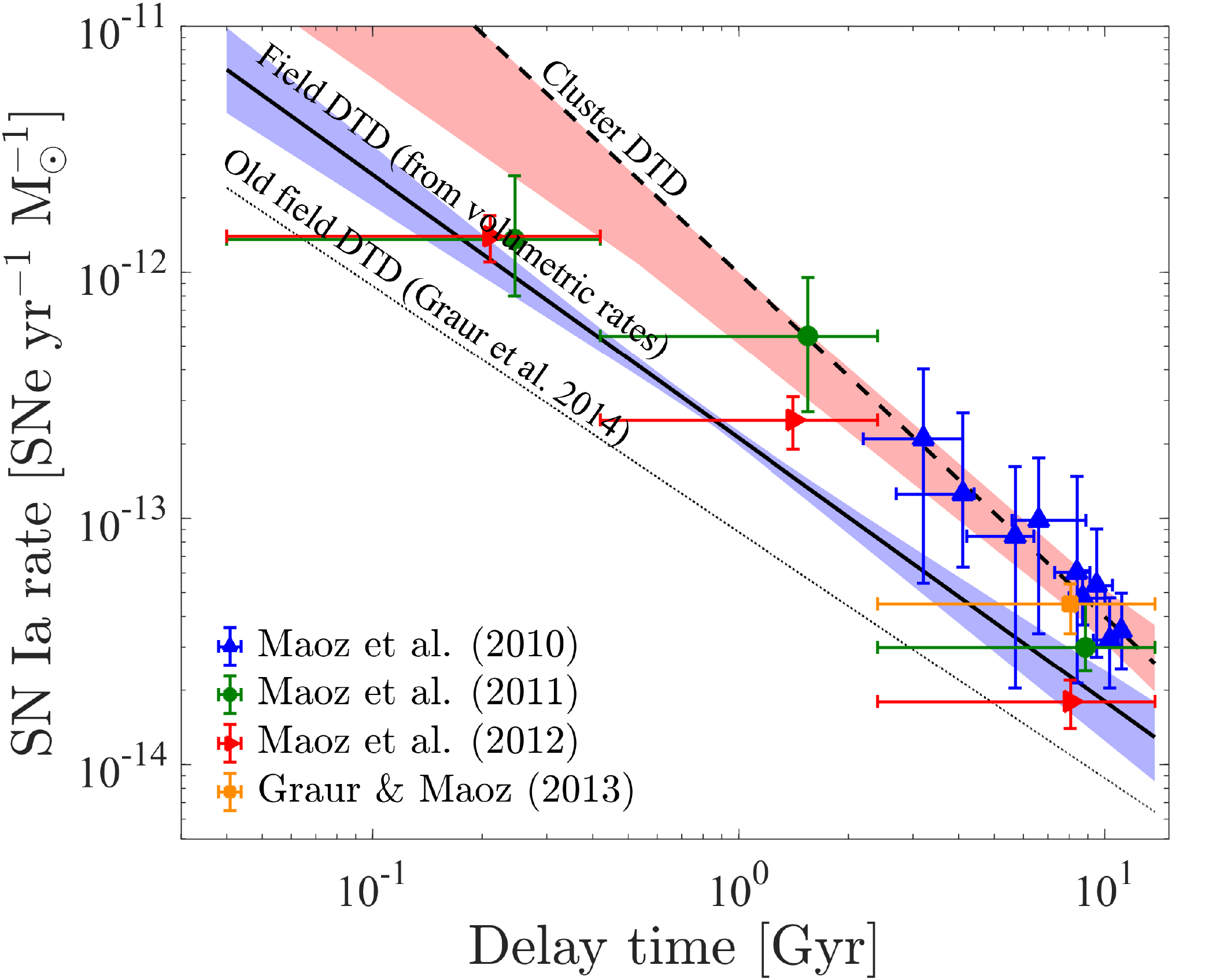}
 \caption{Best-fitting field DTD (solid curve with blue uncertainty bowtie) that, when convolved with the MF17 SFH, matches the SN~Ia rate evolution, as shown in Fig.~\ref{fig:rates}. The DTD previously obtained by \citet{Graur2014} based on the same SN~Ia volumetric rates (Fig.~\ref{fig:rates}), but using the factor $\sim 2$ higher SFH of \citet{2008ApJ...683L...5Y}, is shown with a dotted curve.  Previous DTDs based on field-galaxy SN surveys are also shown, as well the higher-normalization DTD (blue points) obtained from SN Ia rates in galaxy clusters \citep{Maoz2010clusters}. Vertical error bars denote $1\sigma$ errors, and horizontal error bars show the time bin of each measurement. The best-fitting galaxy-cluster DTD, fitted to the higher-normalization DTD measurements, is shown as a dashed curve with a red uncertainty bowtie.} 
\label{fig:dtd}
\end{figure}

Figure~\ref{fig:dtd} shows this best-fitting DTD, along with DTD measurements from other samples and methods, as decribed above and listed in Table~\ref{table:dtd}. Also shown is the best-fitting DTD previously obtained by \citet{Graur2014} based on the same SN~Ia volumetric rates, but using the factor $\sim 2$ higher SFH of \citet{2008ApJ...683L...5Y}. The lower SFH of MF17 naturally requires convolution with a higher-normalized DTD in order to match the same observed evolution of the SN~Ia rate (Fig.\ref{fig:rates}). As seen in Figure~\ref{fig:dtd}, all of the field-galaxy-based DTD determinations are now consistent in both shape and normalization. Considering both the individual uncertainties from each determination and the spread among the different methods, we estimate a power-law index $\alpha=-1.13 \pm 0.06$ and a Hubble-time-integrated SN~Ia production efficiency of $N/M_\star=1.6 \pm 0.3$ SNe per $1000~{\rm M_\odot}$ of stellar mass formed ($\chi^2=10$ for $5$ degrees of freedom). This value is consistent with that from the best-fit DTD to the volumetric SN Ia rates in Figure~\ref{fig:rates}: $N/M_\star=1.3 \pm 0.1$ SNe per $1000~{\rm M_\odot}$ of stellar mass formed. The volumetric-rate-based DTD normalization is actually almost identical to the one from \citet{Maoz2012sdss}, $N/M_\star = (1.30 \pm 0.15)\times10^{-3}~{\rm M}_\odot^{-1}$, which is the most reliable among the DTDs based on individual galaxy SFHs. As already noted, the galaxy-cluster DTD has a higher normalization, of $N/M_\star=(3.3$--$8.1)\times 10^{-3}~{\rm M_\odot}^{-1}$ (see \citealt{Maoz2010clusters}). Using the observed SN-rate-based points in clusters and the normalization constraint, the cluster DTD's power-law index is $\alpha=-1.39^{+0.32}_{-0.05}$. The derived DTD parameters are summarized in Table~\ref{table:NM}.

\citet{2015A&A...584A..62C} have recently studied the observational constraints on the SN~Ia DTD by comparing largely the same set of volumetric SN~Ia rates that we use, and the similar SFH of MD14. However, \citet{2015A&A...584A..62C} fit the rates data with only three particular semi-analytic DTD models of \citet{2005A&A...441.1055G}, rather than to a parametrized power-law model, as we do here. Furthermore, they do not present uncertainties on their best-fit DTD normalizations, nor do they compare their results to any previous DTD determinations.

\begin{table}
 \centering
 \caption{Volumetric SN Ia rates used in this work.}\label{table:rates}
 \begin{tabular}{ccl}
  \hline
  \hline
  {Redshift} & {Rate} & {Source} \\
  {} & {$(10^{-4}~{\rm yr^{-1}~Mpc^{-3}})$} & {} \\
  \hline
  0      & $0.265^{+0.034,+0.043}_{-0.033,-0.043}$ & \citet{li2011rates} \\
  0.0375 & $0.278^{+0.112,+0.015}_{-0.083,-0.000}$ & \citet{dilday2010a} \\
  0.1    & $0.259^{+0.052,+0.018}_{-0.044,-0.001}$ & \citet{dilday2010a} \\ 
  0.11   & $0.247^{+0.029,+0.016}_{-0.026,-0.031}$ & \citet{GraurMaoz2013} \\
  0.15   & $0.307^{+0.038,+0.035}_{-0.034,-0.005}$ & \citet{dilday2010a} \\
  0.2    & $0.348^{+0.032,+0.082}_{-0.030,-0.007}$ & \citet{dilday2010a} \\
  0.26   & $0.32^{+0.08,+0.07}_{-0.08,-0.08}$      & \citet{perrett2012} \\
  0.35   & $0.41^{+0.07,+0.07}_{-0.07,-0.07}$      & \citet{perrett2012} \\
  0.45   & $0.41^{+0.07,+0.05}_{-0.07,-0.06}$      & \citet{perrett2012} \\
  0.55   & $0.55^{+0.07,+0.05}_{-0.07,-0.06}$      & \citet{perrett2012} \\
  0.65   & $0.55^{+0.06,+0.05}_{-0.06,-0.07}$      & \citet{perrett2012} \\
  0.75   & $0.67^{+0.07,+0.06}_{-0.07,-0.08}$      & \citet{perrett2012} \\
  0.85   & $0.66^{+0.06,+0.07}_{-0.06,-0.08}$      & \citet{perrett2012} \\
  0.95   & $0.89^{+0.09,+0.12}_{-0.09,-0.14}$      & \citet{perrett2012} \\
  1.05   & $0.85^{+0.14,+0.12}_{-0.14,-0.15}$      & \citet{perrett2012} \\
  1.23   & $0.84^{+0.25}_{-0.28}$                  & \citet{Graur2011} \\
  1.59   & $0.45^{+0.34,+0.05}_{-0.22,-0.09}$      & \citet{Graur2014} \\
  1.61   & $0.42^{+0.39,+0.19}_{-0.23,-0.14}$      & \citet{dahlen2008} \\
  1.69   & $1.02^{+0.54}_{-0.37}$                  & \citet{Graur2011} \\
  1.75   & $0.71^{+0.45,+0.50}_{-0.29,-0.30}$      & \citet{2014AJ....148...13R} \\
  2.25   & $0.55^{+0.97,+0.53}_{-0.41,-0.29}$      & \citet{2014AJ....148...13R} \\
  \hline
  \multicolumn{3}{l}{Note. Statistical uncertainties are followed by systematic} \\
  \multicolumn{3}{l}{uncertainties, separated by commas.}
 \end{tabular}
\end{table}

\begin{table}
\center
\caption{SN Ia DTD measurements used in this work.}\label{table:dtd}
 \begin{tabular}{ccl}
  \hline
  \hline
  {Delay} & {DTD} & {Source} \\
  {(Gyr)} & {$(10^{-14}~{\rm yr^{-1}~M_\odot^{-1}})$} & {} \\
  \hline
  \multicolumn{3}{c}{Field galaxies} \\
  \hline
  $0.21^{+0.21}_{-0.17}$ & $136^{+110}_{-56}$             & \citet{Maoz2010loss} \\
  $0.21^{+0.21}_{-0.17}$ & $140^{+30}_{-30}$              & \citet{Maoz2012sdss} \\
  $1.4^{+1.0}_{-1.0}$    & $55^{+40}_{-28}$               & \citet{Maoz2010loss} \\
  $1.4^{+1.0}_{-1.0}$    & $25^{+6}_{-6}$                 & \citet{Maoz2012sdss} \\
  $8.1^{+5.7}_{-5.7}$    &  $3^{+1.5}_{-0.6}$             & \citet{Maoz2010loss} \\
  $8.1^{+5.7}_{-5.7}$    &  $1.8^{+0.4}_{-0.4}$           & \citet{Maoz2012sdss} \\
  $8.1^{+5.7}_{-5.7}$    &  $4.5^{+0.6,+0.3}_{-0.6,-0.5}$ & \citet{GraurMaoz2013}$^{a}$ \\
  \hline
  \multicolumn{3}{c}{Galaxy clusters} \\
  \hline
  $3.1^{+0.9}_{-1.0}$  & $20^{+18}_{-15}$ & \citet{Maoz2010clusters} \\
  $4.1^{+0.3}_{-1.4}$  & $12^{+13}_{-6}$  & \citet{Maoz2010clusters} \\
  $5.7^{+0.7}_{-1.5}$  &  $8.0^{+7.2}_{-6.1}$   & \citet{Maoz2010clusters} \\
  $6.6^{+2.3}_{-1.0}$  &  $9.3^{+7.2}_{-6.1}$   & \citet{Maoz2010clusters} \\
  $8.4^{+0.7}_{-1.1}$  &  $5.7^{+8.3}_{-3.7}$   & \citet{Maoz2010clusters} \\
  $8.7^{+1.3}_{-0.7}$  &  $4.5^{+1.3}_{-1.0}$   & \citet{Maoz2010clusters} \\
  $9.5^{+1.0}_{-0.4}$  &  $5.0^{+3.4}_{-2.5}$   & \citet{Maoz2010clusters} \\
  $10.3^{+0.6}_{-1.0}$ &  $3.0^{+1.5}_{-1.1}$   & \citet{Maoz2010clusters} \\
  $11.1^{+0.2}_{-0.3}$ &  $3.3^{+1.4}_{-1.0}$   & \citet{Maoz2010clusters} \\
  \hline 
  \multicolumn{3}{l}{Note. DTD values in galaxy clusters from \citet{Maoz2010clusters}}\\
  \multicolumn{3}{l}{have been reduced by 9\% to convert from the \citet{2003ApJS..149..289B}}\\
    \multicolumn{3}{l}{``diet'' Salpeter IMF used by \citet{Maoz2010clusters} to}\\
  \multicolumn{3}{l}{the \citet{2001MNRAS.322..231K} IMF consistently used here.} \\
  \multicolumn{3}{l}{$^a$Systematic uncertainties separated by comma from statistical} \\
  \multicolumn{3}{l}{uncertainties.} \\
 \end{tabular}
\end{table}

\begin{table}
 \centering
 \caption{SN Ia DTD parameters.}\label{table:NM}
 \begin{tabular}{lccc}
 \hline
 \hline
 {Type} & {$t^\alpha$} & {$N/M_\star$} & {$\chi^2/{\rm DOF}$} \\
 {} & {} & {$(\times10^{-3}~{\rm M_\odot})$} & {} \\
 \hline
 Field galaxies   & $-1.13^{+0.06}_{-0.06}$ & $1.6^{+0.3}_{-0.3}$ & $10/5$ \\
 Volumetric rates & $-1.07^{+0.09}_{-0.08}$ & $1.3^{+0.1}_{-0.1}$ & $7/19$ \\
 Galaxy clusters  & $-1.39^{+0.32}_{-0.05}$ & $5.4^{+2.3}_{-2.3}$ & $0.6/7$ \\
 \hline
 \multicolumn{4}{l}{Note. All $N/M_\star$ values consistently assume the} \\
 \multicolumn{4}{l}{\citet{2001MNRAS.322..231K} IMF.} \\
 \end{tabular}
\end{table}


\section{Iron and $\alpha$-element buildup history}
\label{sec:iron}

Iron is produced and dispersed solely by SNe~Ia and CC~SNe. The predominant stable isotope of iron, $^{56}$Fe, is a beta-decay product of the $^{56}$Ni synthesized in SNe, through its daughter nucleus $^{56}$Co. The energy release from this decay chain is the main driver of SN~Ia light curve luminosity (see \citealt{2016ApJ...819...31G} regarding the dominance of the $^{57}$Co decay chain at late times), but plays a significant energetic role in many CC~SN light curves as well. The final $^{56}$Fe yield can therefore be estimated for both types of SN light curves. The yield in other iron isotopes can be estimated from detailed spectroscopic modeling of SN events. There is thus a good semi-empirical handle on the total iron yield of different SN types. 

The CC~SN rate versus redshift, as already noted, is just the SFH multiplied by the number of stars which explode as CC~SNe, per formed unit of stellar mass. The integrated CC~SN rate history times the mean CC~SN iron yield per explosion gives the CC~SN contribution to the cosmic iron budget. If the SN~Ia DTD is known, it can be convolved with the cosmic SFH to give the volumetric SN~Ia rate at all redshifts (including redshifts beyond those where the SN~Ia rate has been measured directly). As with the CC~SNe, the SN~Ia rate times the mean iron yield of SNe~Ia, integrated over cosmic time, gives the accumulation history of volumetric iron mass density from SNe~Ia. 

\citet{Graur2011} performed the above estimate of the iron accumulation history, using previous SFH estimates, a SN~Ia DTD obtained from the old (factor-about-2 high) SFH, and some rough mean iron yields for SNe~Ia and CC~SNe. We now repeat this exercise with the revised SFH of MF17, the resulting revised SN~Ia DTD, and using improved empirical estimates of CC~SN iron yields. 

The mean iron yield of a SN~Ia (the mean yield is all that is required in this kind of estimate, under the assumption that the yield distribution of SNe~Ia does not evolve, e.g., due to dependence of
yield on progenitor metallicity) is (e.g., \citealt{2007Sci...315..825M,Howell2009}) 
\begin{equation}\label{eq:eq2}
 \bar y_{\rm Fe, Ia}=0.7~{\rm M_\odot}, 
\end{equation}
as already assumed in \citet{Graur2011}. For CC~SN iron yields, we use the compilation of \citet{2015arXiv150602655K}, from which we find that Type IIP SNe have a median iron yield of $0.03~{\rm M_\odot}$
(consistent with \citealt{2017ApJ...841..127M}), while Type Ib and Ic SNe have a median iron yield of $0.2~{\rm M_\odot}$. SNe IIb and SN 1987A-like SNe have median yields of $0.12$ and $0.08~{\rm M_\odot}$, respectively. From the local volume-limited LOSS sample of SNe (\citealt{li2011LF,2017ApJ...837..120G,2017ApJ...837..121G,2017PASP..129e4201S}), $\sim70$\% of CC~SNe are Type II and $\sim30$\% are stripped-envelope SNe, of which $\sim36$\% are SNe IIb and the rest are SNe Ib, Ic, and their peculiar variants. Among the Type II SNe, $\sim90$\% are Type IIP or IIL, $\sim4$\% are SN 1987A-like SNe, and the rest are SNe IIn (for simplicity, we will assume that the latter have similar iron yields as SNe IIP). Assuming also that the mix of CC~SN types (and their iron yields) in LOSS is representative of the general field population, and that this mix does not evolve, the mean CC~SN iron yield is
\begin{equation}\label{eq:eq3}
 \begin{split}
  \bar y_{\rm Fe, cc} & = (0.7\times0.96)\times 0.03~{\rm M_\odot} + (0.7\times0.04)\times0.08~{\rm M_\odot} \\
  & + (0.3\times0.36)\times0.12~{\rm M_\odot} + (0.3\times0.64) \times 0.2~{\rm M_\odot} \\
  & = 0.074~{\rm M_\odot}.  
 \end{split}
\end{equation}
By comparison, \citet{Graur2011} used values of either $\bar y_{\rm
  Fe,CC}=0.066~{\rm M_\odot}$ (following \citet{2008NewA...13..606B}),
or 1\% of the CC SN progenitor mass (following
\citet{Maoz2010clusters}). Our adopted yield in intermediate to these two. 
 
\begin{figure}
 \centering
 \includegraphics[width=0.47\textwidth]{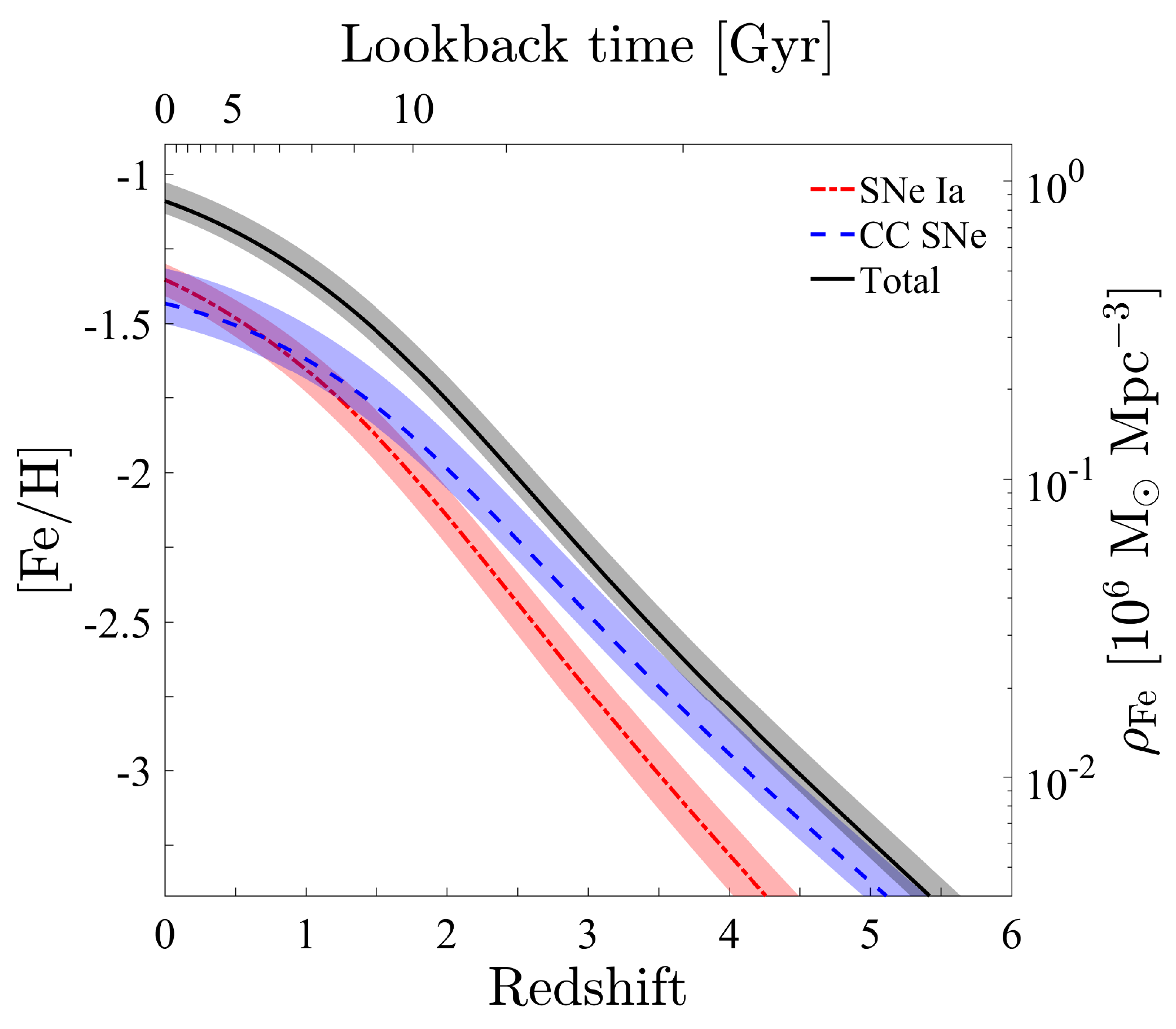}
 \caption{Cosmic mean iron volume density (right-hand axis), and iron abundance, relative to Solar, [Fe/H] (left-hand axis), versus redshift, contributed by CC~SNe (based on the SFH of MF17), by SNe~Ia (based on the best-fitting DTD from Figure~\ref{fig:rates} convolved with the SFH), and by their sum. The bands around the SN Ia, CC SN, and summed curves represent, respectively, the 68\% uncertainty region of the SN Ia fit, the uncertainty on the minimal mass of CC SN progenitors, $8.5^{+1.5}_{-1.0}~{\rm M_\odot}$ \citep{Smartt2009review}, and the combination of these uncertainties, squared. At the present epoch, SNe~Ia and and CC~SNe have made similar contributions to iron production, with a slight dominance of SNe~Ia since $z\approx0.7$.}
 \label{fig:iron}
\end{figure}

\begin{figure}
 \centering
 \includegraphics[width=0.47\textwidth]{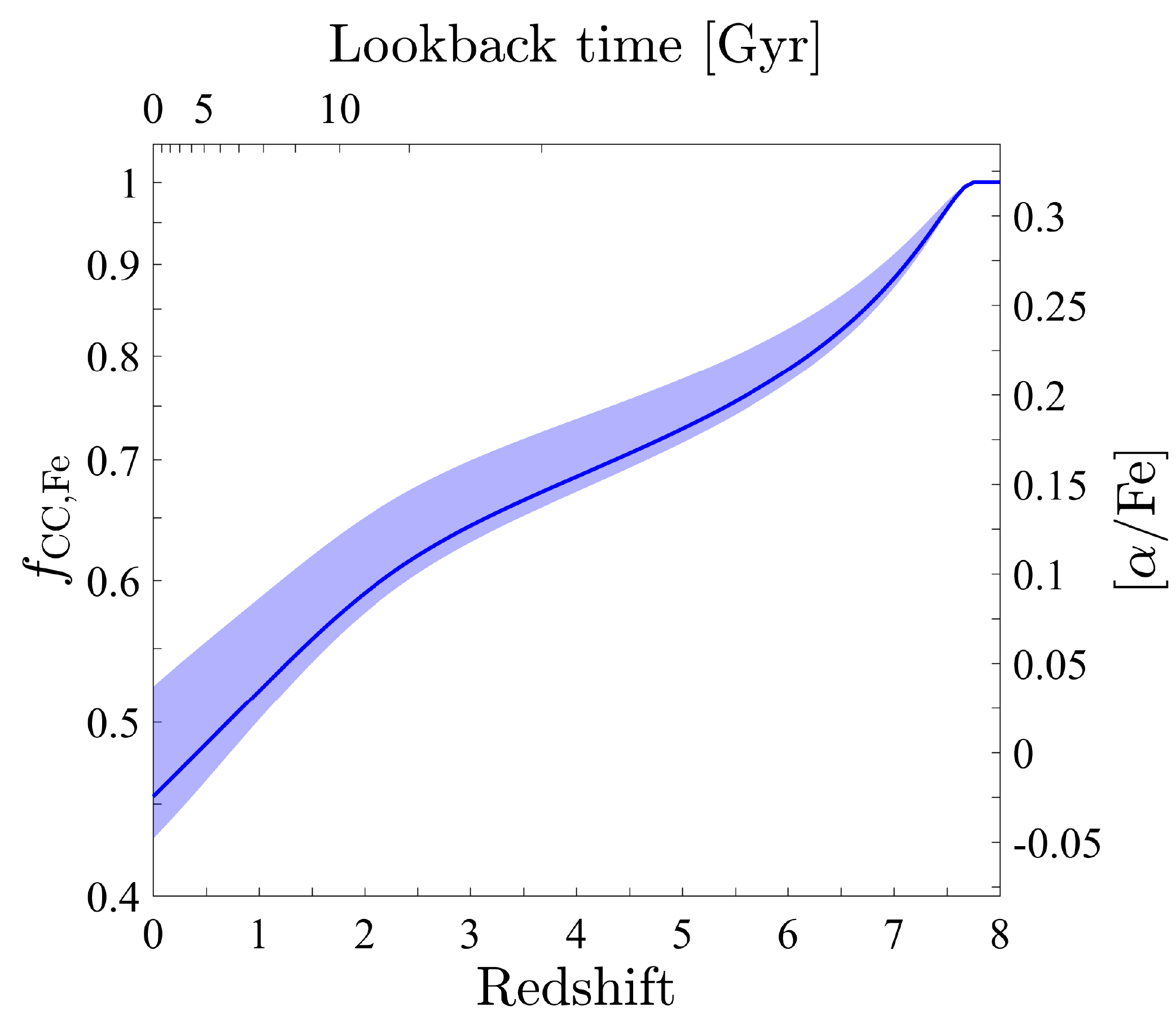}
 \caption{Relative CC~SN contribution (left-hand axis) to the total iron budget (from both CC~SNe and SNe~Ia) as a function of redshift. Right-hand axis translates the dependent variable to the $[{\alpha}/{\rm Fe}]$ ratio 
of a mean cosmic volume element, by means of Eq.~\ref{eq:alphaFe}.}
 \label{fig:fCCz}
\end{figure}

Figure~\ref{fig:iron} shows the volumetric iron accumulation history versus redshift, from SNe~Ia, from CC~SNe, and from their sum. As in \citet{Graur2011}, the volumetric iron-mass density, $\rho_{{\rm tot, Fe}}(z)$, can be divided by the comoving baryon density ($\rho_{\rm baryon}$, with values from \citealt{2011ApJS..192...18K}) times the Solar iron mass abundance ($Z_{\rm Fe,\odot}=0.00174$; \citealt{2009ARA&A..47..481A}), to express the iron accumulation history in terms of the mean Fe abundance, relative to Solar, 
\begin{equation}
\label{eq:FeH}
\left[\frac{\rm Fe}{\rm H}\right](z)=\log\frac{\rho_{{\rm tot, Fe}}(z)}{\rho_{{\rm baryon}}Z_{\rm Fe,\odot}}.
\end{equation}
This is shown on the left-hand axis of Fig.~\ref{fig:iron}. As opposed to the result in \citet{Graur2011}, which indicated dominance of CC~SN contributions to the present-day iron abundance, our new estimate, based on the scaled-down SFH of MF17, shows a similar contribution to iron production by SNe~Ia and by CC~SNe, with a slight dominance by SNe~Ia ($55 \pm 6$\% of the total, with the SNe~Ia having become dominant since $z\approx0.7$). From the summed SN contribution, the present-day mean logarithmic cosmic iron abundance, relative to Solar, is ${\rm [Fe/H]} = -1.09^{+0.07}_{-0.04}$ (i.e., $\sim 8$\% of Solar). Figure~\ref{fig:fCCz} shows (on its left-hand axis) $f_{\rm CC, Fe}$, the accumulated fractional mass contribution by CC~SNe to the total iron mass budget.

These empirically derived relative contributions of SNe~Ia and CC~SNe to the iron budget can also be used to illuminate the cosmic history of $\alpha$-element abundance. $\alpha$ elements are produced
predominantly by CC~SNe (see, e.g., \citealt{2017ApJ...837..183W}). For example, a typical SN~Ia makes about $0.01~{\rm M_\odot}$ of calcium \citep{2013MNRAS.429.1156S}, while a CC~SN makes of order $0.05$--$0.1~{\rm M_\odot}$ of calcium \citep{2013ARA&A..51..457N}, and, integrated over time, there are about about seven or eight CC~SNe for every SN~Ia (compare $N/M_\star$ of 0.01~${\rm M_\sun^{-1}}$ for CC~SNe, as already noted, to $N/M_\star \approx 0.0013~{\rm M_\sun^{-1}}$ for SNe Ia, as we have found). Thus, no more than a few per cent of all calcium has been produced in SNe~Ia. We will therefore neglect the SN~Ia contribution to $\alpha$-element production. In contrast, iron, as noted, is made by both CC~SNe and SNe~Ia. Consider the volume of gas from which the Sun was formed, containing at the time of Solar formation a mass $M_{\alpha,\odot}$ of $\alpha$ elements. Let us denote with $M_{\alpha}(z)$ the $\alpha$-element mass in this same gas volume at the time corresponding to a pre-Solar-formation redshift of $z$. Further denoting as $N_{{\rm CC},\odot}$ and $N_{{\rm CC}}(z)$ the number of CC~SNe that had enriched the gas volume by the time of the Sun's formation and by redshift $z$, respectively, then 
\begin{equation}
 \frac{M_{\alpha}(z)}{M_{\alpha,\odot}}=\frac{N_{{\rm CC}}(z)}{N_{{\rm CC},\odot}}.
\end{equation}
Multiplying the right-hand-side numerator and denominator by the mean iron 
yield of CC~SNe, $\bar y_{\rm Fe, CC}$, we see that
\begin{equation}
 \frac{M_{\alpha}(z)}{M_{\alpha,\odot}}=\frac{M_{{\rm CC, Fe}}(z)}{M_{{\rm CC,Fe},\odot}}.
\end{equation}
In other words, the $\alpha$ abundance of the cloud at a redshift $z$, relative to Solar, equals the CC~SN-contributed iron abundance, relative to Solar. Dividing the numerators on both sides of the equation by the iron mass in the cloud (from both CC~SNe and SNe~Ia) at redshift $z$, and both denominators by the iron mass in the Sun, and taking the logarithm, we get the $\alpha$ to iron abundance in this pre-Solar cloud, relative to Solar, at redshift $z$:
\begin{equation}
\label{eqalphaFe1}
 \left[\frac{\alpha}{\rm Fe}\right](z)=\log\frac{f_{{\rm CC, Fe}}(z)}{f_{{\rm CC,Fe},\odot}},
\end{equation}
where we recall that $f_{\rm CC, Fe}$ is the fractional mass contribution by CC~SNe to the total iron budget. For some star that formed from this gas at a very early time, after some CC~SN enrichment of the gas had taken place but no SN~Ia explosion had yet enriched the gas, $f_{\rm CC,Fe}(z)=1$, and $[{\alpha}/{\rm Fe}](z)=-\log{f_{{\rm CC,Fe},\odot}}$. Thus, the ``plateau'' value of $[{\alpha}/{\rm Fe}]\approx 0.3$ to $0.5$, observed in the metal-poorest stars in the MW, indicates that CC~SNe have contributed no more than 50\% of the Sun's iron.

As seen in Figure~\ref{fig:fCCz}, at redshifts below $z\approx 1$, $f_{\rm CC, Fe}(z)=0.45$ to $0.55$, as expected from the $[{\alpha}/{\rm Fe}]$ ratio observed in metal poor stars. This kind of $f_{\rm CC, Fe}$ is deduced for the Sun (see above) and is expected in any environment that has undergone enrichment for a sufficient amount of time, and simply reflects the time-integrated iron yields of SNe~Ia and CC~SNe per unit formed stellar mass. At $z=0.43$, corresponding to a lookback time of 4.5~Gyr (the Sun's age), $f_{\rm CC, Fe}(z)=0.48^{+0.07}_{-0.02}$. The agreement between the low-$z$ cosmic value of the CC~SN fractional contribution to the iron budget, $f_{\rm CC,Fe}$, and the $[{\alpha}/{\rm Fe}]$ ratio of metal-poor stars, did not exist when using the previous iron enrichment history of \citet{Graur2011}, which was based on overall higher SFHs, and therefore led to a CC~SN dominance (higher $f_{\rm CC, Fe}$). This is another paradox that is resolved by adopting the down-revised cosmic SFH. Replacing, in Eq.~\ref{eqalphaFe1}, $f_{{\rm CC,Fe},\odot}$ with $f_{{\rm CC,Fe}}(z=0.43)$, we can directly translate the cosmic history of the fractional CC-SN contribution to the iron budget, $f_{{\rm CC, Fe}}(z)$, into a cosmic history of the mean $[{\alpha}/{\rm Fe}]$ ratio,
\begin{equation}\label{eq:alphaFe}
 \left[\frac{\alpha}{\rm Fe}\right](z)=\log\frac{f_{{\rm CC, Fe}}(z)}{f_{{\rm CC,Fe}}(z=0.43)},
\end{equation}
which is indicated on the right-hand axis of  Fig.~\ref{fig:fCCz}.

\begin{figure*}
 \centering
 \begin{tabular}{cc}
  \includegraphics[width=0.47\textwidth]{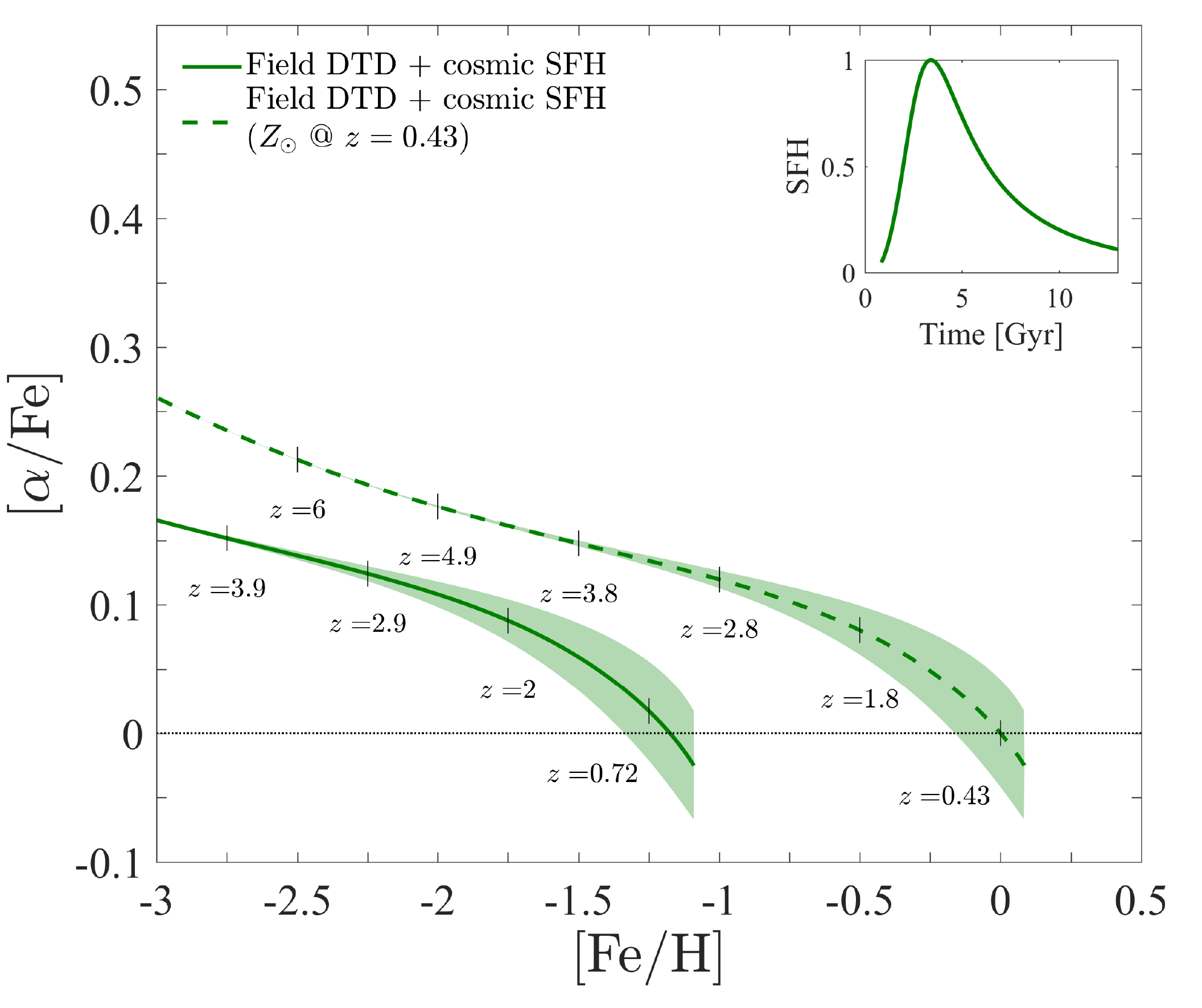} & 
  \includegraphics[width=0.47\textwidth]{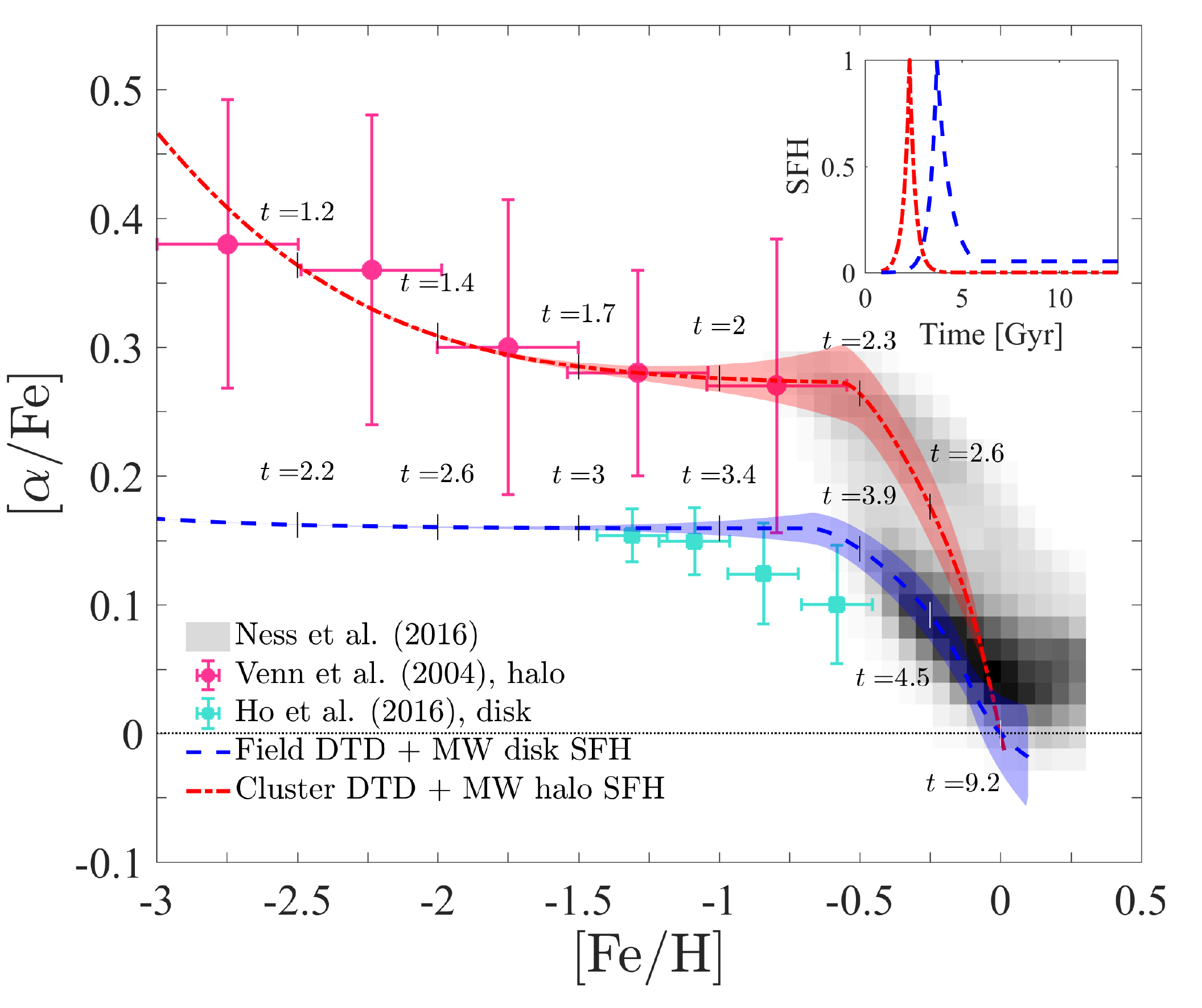}
 \end{tabular}
 \caption{{\it Left:} Cosmic $[{\alpha}/{\rm Fe}]$ to $[{\rm Fe/H}]$ relation (green solid curve) for a mean volume element in the Universe, plotted through Eqns.~\ref{eq:FeH} and \ref{eq:alphaFe}, and based solely on the observed SFH of MF17 (shown in inset), on the observationally derived field-galaxy SN~Ia DTD from Figure~\ref{fig:dtd}, and on the empirical iron yields of SNe. The green dashed curve shows the evolution of a hypothetical volume element that was enriched by a scaled-up version of the MF17 SFH, such that the gas reaches Solar metallicity 4.5~Gyr ago (Eqns.~\ref{eq:alphaFe} and \ref{eq:FeH2}). Line thickness (semi-transparent) is proportional to the total stellar mass produced at each point along the track. Tick marks denote the redshifts along the curve.
 {\it Right:} Observed abundance data for MW stars are shown as error bars and a grey-scale map. Measurements for halo stars (from the compilation of \citealt{2004AJ....128.1177V}) and for thin-disk stars
 \citep{2017ApJ...836....5H} are binned, and the bin medians are shown as filled magenta circles and cyan squares, respectively. Vertical error bars give the standard deviation and horizontal error bars the
 bin width. Higher-metallicity stellar measurements from \citet{2016ApJ...823..114N} are shown as a density map (darker pixels contain more stars). The curves are our calculated MW abundance tracks, obtained by combining: (1; lower track) a simple MW-thin-disk SFH (shown in inset), scaled so as to reach Solar metallicity 4.5~Gyr ago,  with the field galaxy SN~Ia DTD and the empirical iron yields of SNe; and (2; upper track) same, but with a brief $z=3$ burst SFH (also shown in inset) with a galaxy-cluster-like SN~Ia DTD (Figure~\ref{fig:dtd}). Tick marks along the curves denote time since the Big Bang, in Gyr, as in the horizontal axes of the inset SFH plots. These two SFH$+$DTD combinations reproduce well the two main observed stellar abundance loci.}
 \label{fig:fig5}
\end{figure*}

Going one step further, in Figure~\ref{fig:fig5} we combine Eqns.~\ref{eq:FeH} and \ref{eq:alphaFe} in a parametric representation (with $z$ as the parameter) to plot the evolution of the $[{\alpha}/{\rm Fe}]$ ratio as a function of [Fe/H] abundance for the mean cosmic gas volume. Another interesting option is to replace the iron abundance history of  Eq.~\ref{eq:FeH} with  \begin{equation}\label{eq:FeH2}
\left[\frac{\rm Fe}{\rm H}\right](z)=\log\frac{\rho_{{\rm tot, Fe}}(z)}{\rho_{{\rm tot, Fe}}(z=0.43)},
\end{equation}
i.e., to examine the iron abundance not relative to its present-day mean cosmic abundance, but rather relative to the abundance at the time of the Sun's formation. This function, also shown in Figure~\ref{fig:fig5}, describes the elemental evolution of a volume of gas that underwent enrichment by a SFH with a redshift dependence like the MF17 cosmic SFH, but scaled up such that the gas achieved
Solar metallicity by $z=0.43$. Since these descriptions of the relative cosmic iron abundance relate abundance directly to redshift and hence to look-back time, we can also translate [Fe/H] abundance to
stellar age. We note that, as a result of our valid approximation that $\alpha$ elements come solely from CC~SNe, and the fact that we express all abundances relative to Solar, we can make this calculation
using only iron yields, without the need to know or assume any $\alpha$-element SN yields (see also \cite{2017ApJ...837..183W}). The plotted thickness of the cosmic abundance track reflects the stellar mass produced at each point along the track, according to the SFR times the time-interval corresponding to each abundance interval. 

The cosmic $[\alpha/{\rm Fe}]$ to [Fe/H] relations plotted in Figure~\ref{fig:fig5} consist of a shalow decline in $[\alpha/{\rm Fe}]$ when going to higher [Fe/H], up to a ``knee'' at [Fe/H]$\sim -0.5$ (or at [Fe/H]$\sim -1.5$ in the cosmic mean curve), beyond which the fall steepens toward the Solar abundance point at (0,0) (or toward the present-day cosmic abundance in the cosmic mean curve). This behavior is broadly similar in form to that of the two main loci of points that emerge when MW stars are plotted in this abundance-parameter space, as further discussed below. In the MW context, it has often been stated (e.g., \citealt{1986A&A...154..279M}) that the knee is the result of the beginning of iron input from SNe~Ia that ``kick in'' after a long delay relative to the SFR and to the CC~SNe. From Figs~\ref{fig:rates}, \ref{fig:iron}, and \ref{fig:fCCz}, however, it is clear that this is not the case, at least not for these cosmic abundance tracks. The knee occurs roughly at $z\approx 1.5$--$2$, i.e., 3--4~Gyr after the Big Bang. This time is much longer than the initial and brief, $\sim 40$~Myr, delay of the SN~Ia DTD, afer which the DTD jumps to its maximum and begins its power-law decline. During those first few Gyrs, the SN~Ia rate more-or-less tracks the CC SN rate, leading to the shallow trajectory in the abundance-parameter plane. At later times, however, the SFH and the CC SN rate decline steeply, while the SN~Ia rate, which is a convolution of the SFH with the $\sim t^{-1}$ DTD, declines more gradually. The knee is thus actually the result of the CC SNe dying out (or ``kicking the bucket'') rather than the SNe~Ia kicking in. 

In view of the similarity between the cosmic evolution tracks that we have derived in the $[{\alpha}/{\rm Fe}]$ versus $[{\rm Fe/H}]$ plane and the observed stellar loci in that plane, we now briefly investigate what insights we can gain regarding the physical processes behind the latter. Figure~\ref{fig:fig5} shows, in this parameter plane, the observed MW stellar abundances from the Kepler, APOGEE, and LAMOST surveys \citep{2016ApJ...823..114N,2017ApJ...836....5H} and from the compilation of halo-star abundances presented by \citet{2004AJ....128.1177V}. For the $[\alpha/{\rm Fe}]$ of each star, we plot the averages of some or all of the $\alpha$ elements O, Mg, Si, Ca, S, and Ti, as computed by the above authors. It has been noted for some time that, for [Fe/H]$\gtrsim -1.0$, stars are concentrated in this plane in two sequences, a ``high-$\alpha$'' and a ``low-$\alpha$'' sequence (e.g., \citealt{2004AN....325....3F,2009MNRAS.399.1145S,2012ApJ...753..148B}). The median [$\alpha$/Fe] values of the halo stars of \citet{2004AJ....128.1177V} in bins of width 0.5 in [Fe/H], as plotted in Fig.~\ref{fig:fig5}, appear to extend, toward lower metallicities, the high-$\alpha$ branch seen in the data of \citet{2016ApJ...823..114N}. The points based on \citet{2017ApJ...836....5H}, consisting of the medians and standard deviations within the plotted bins of the stars with signal-to-noise ratios $>20$, [Fe/H]$<-0.5$, and [$\alpha$/Fe]$<0.17$ (the rough borders of the low-$\alpha$ locus visible in these data), roughly match onto the low-$\alpha$ branch of \citet{2016ApJ...823..114N}. These most recent precision-stellar-abundance data shown bring these two loci into new focus. 

There is yet no consensus in the literature on the exact form, nature, or origin of the observed trends of stars in the abundance-parameter plane. Traditionally, the low-$\alpha$ sequence has been associated with the MW thin disk, and the high-$\alpha$ track with the thick disk and the halo (\citealt{2006MNRAS.367.1329R,2011MNRAS.414.2893F,2011MNRAS.412.1203N}). \citet{2012ApJ...753..148B}, however, claimed that there is no distinct thick disk component but rather a continuum of components with a range of scale heights and scale radii, and that the apparent bimodality is an artifact that disappears when sample selection is properly corrected for. \citet{2016A&A...589A..66H} prefer to associate the two $\alpha$ sequences with an ``inner disk'' and an ``outer disk,'' rather than with thick- and thin-disk components, and argue that stellar abundance tracks cross the gap between the two sequences, with the gap resulting from a temporary shutoff in star formation 8 Gyr ago. \citet{2016A&A...594A..61W} have noted a strong similarity between the abundance patterns in MW stars and in the stellar populations of early-type galaxies, and concluded that a power-law DTD, as we use here, is required in order to reproduce those patterns. \citet{2017ApJ...836....5H} reaffirm the reality of a low-$\alpha$ sequence concentrated in the Galaxy's midplane, and of an old, high-$\alpha$, population associated with the thick disk, the halo, and the outer bulge of the Galaxy.
 
The resemblance between the abundance evolution track that we have derived for a gas element enriched by a cosmic-like SFH, but reaching Solar metallicity 4.5~Gyrs ago, and the observed low-$\alpha$ MW stellar locus, suggests that the cosmic SFH is not too different, in its general form, from the MW disk's ``effective'' SFH, i.e., the SFH that would create the same metal enrichment in a chemically evolving, closed-box, gas volume, as would a SFH in a calculation that includes effects such as dilution by inflow of pristine gas, loss to outflows of enriched gas, and recycling of unenriched gas from stellar winds. Some past attempts at reconstructing the Galaxy's SFH using stellar abundances (e.g., \citealt{2016A&A...589A..66H}) indicate such a resemblance, in the form of vigorous star formation for the first few Gyr, but then a steep, order-of-magnitude decline in the MW star-formation rate starting 10 Gyr ago, corresponding to $z\approx 2$, i.e., not very different from the general form of the cosmic SFH. 

\citet{2017ApJ...837..183W} and \citet{2017ApJ...835..224A} (and to some degree also \citealt{2017arXiv170208729R}) have recently used relatively simple chemical evolution models to explore how the various input parameters of their models affect the evolution tracks of gas in the abundance parameter space. Following these leads, we have experimented with varying only the effective SFH in an even-simpler model, to see if we can easily find a Galactic SFH that matches well the MW-disk distribution of stellar abundances in the $[{\alpha}/{\rm Fe}]$--[Fe/H] plane. For each trial SFH, we use the SN~Ia DTD and the SN iron yields to find the SN~Ia and CC~SN iron enrichment histories, and derive the $[{\alpha}/{\rm Fe}]$ to [Fe/H] relation, as we did for the cosmic SFH case. In the terminology of chemical evolution models, we are calculating the simplest possible ``one-zone, closed-box'' model, with an effective SFH scaled to produce the Solar [Fe/H] and $[{\alpha}/{\rm Fe}]$ abundances at cosmic time $t=9.2$~Gyr (i.e., 4.5 Gyr ago). We recall here that, apart from the field-galaxy-based SN~Ia DTD, there is evidence for a higher-normalized DTD in galaxy clusters (see Section~\ref{sec:DTD} and Fig.~\ref{fig:dtd}), whose SFHs are likely to have been brief single bursts of star formation at $z\sim 2$--$3$ (e.g., \citealt{2010ApJ...720..284M}). We therefore further test if, by assuming such a SFH for the thick-disk/halo component of the MW, and combining it with the a cluster-like SN~Ia DTD, we can reproduce the high-$\alpha$ locus of MW halo stars.
 
Figure~\ref{fig:fig5} shows that, indeed, we can reproduce well both of the main stellar loci in the $[{\alpha}/{\rm Fe}]$--[Fe/H] diagram. The low-$\alpha$ distribution of thin-disk abundances is matched by assuming a MW SFH that, starting at $t_1=0.85$ Gyr (i.e., $z=8$), rises as ${\rm exp}[(t-t_2)/0.35~{\rm Gyr}]$ until $t_2=3.7$ Gyr (10~Gyr ago, or $z=1.85$), and then declines exponentially as ${\rm exp}[-(t-t_2)/0.68~{\rm Gyr}]$ for 2 Gyr (until $z=1.07$). The SFR then levels off at a constant level which is 5\% of the peak level. The ratio between the integral over this assumed SFH and the late constant level, $\sim 30$~Gyr, is consistent with the ratio of the MW disk's stellar mass ($(7.4\pm 1.6)\times 10^{10}~{\rm M_\odot}$, \citealt{2015ApJ...806...96L}), and the current MW star-formation rate of $2.7 \pm 0.3~{\rm M}_\odot {\rm yr}^{-1}$ (\citealt{2011AJ....142..197C}, both corrected to the \citet{2001MNRAS.322..231K} IMF used here).

To match the high-$\alpha$ locus of halo stars, we assume a single-burst SFH that peaks at $t_3=2.3$ Gyr, or $11.4$~Gyr ago (corresponding to $z=3$), of the form ${\rm exp}[(t-t_3)/0.29~{\rm Gyr})]$ before the peak, and ${\rm exp}[-(t-t_3)/0.29~{\rm Gyr})]$ after the peak. As noted, this SFH is convolved with the best-fitting ``galaxy-cluster'' DTD from Table~\ref{table:NM}, with its high normalization of $N/M_\star=5.4\times 10^{-3}$~M$_\odot^{-1}$. As seen in Figure~\ref{fig:fig5}, this results in an excellent match to the high-$\alpha$ track. Intriguingly, it is difficult to find any plausible MW SFH that, when combined with the field-galaxy DTD (rather than with the galaxy-cluster DTD), reproduces the high-$\alpha$ sequence. A SFH that does this roughly is an extremely early ($z\sim 8$) and brief ($\lesssim 100$~Myr) burst that makes all halo and thick-disk stars within that time interval. However, such a SFH would be in conflict with the actual ages derived for such stars (\citealt{2016ApJ...823..114N,2017ApJ...836....5H}; e.g., their figure~8).
  
The effective SFHs we use for the two MW components are also shown in Figure~\ref{fig:fig5}. The widest parts of the plotted calculated abundance tracks, which correspond to the regions with the most stars
formed, roughly line up with the most densely populated parts of the diagram, as would be expected. Our search for MW SFHs that match the observed stellar abundance relations has been by no means exhaustive, and hence we do not claim that the above SFHs are either novel (similar histories for the thin and thick disk components have often been invoked) or that they are unique fits to the data. Rather, our point is that, using the fixed, observationally determined, SN~Ia DTDs and SN iron yields, it is easy to find at least some simple SFHs that, when incorporated into the simplest possible chemical evolution scheme, provide surprisingly good matches to the stellar data. In future work, we will pursue detailed modeling that explores more fully the parameter space of the Galactic SFH, while considering additional observational constraints, such as the measured ages of the individual stars and their detailed densities in the parameter space.

\section{Conclusions}
\label{sec:conclusions}

Numerous SN Ia rate measurements in field galaxies have indicated a universal and consistent DTD---a power law of the form $t^{-1.13\pm0.06}$, with a Hubble-time integrated SN production efficiency of $N/M_\star=(1.6\pm 0.3)\times 10^{-3}$~M$_\odot^{-1}$ (for an assumed \citealt{2001MNRAS.322..231K} IMF). In this work, we have shown that a previously discrepant value of $N/M_\star$ of the DTD, obtained by comparing volumetric SN~Ia rates to the cosmic SFH, falls into line with other measurements once a modern SFH is used in the derivation, particularly a SFH that does not have over-estimated star formation at $z>2$. The new volumetric-rate-based DTD has the form $t^{-1.07\pm0.09}$ and $N/M_\star=(1.3\pm 0.1)\times 10^{-3}$~M$_\odot^{-1}$, fully consistent in both slope and normalization with previous measurements. Apart from this field-galaxy-based DTD, previous studies have suggested the existence of a distinct SN~Ia DTD in galaxy cluster environments, of the form $t^{-1.39^{+0.32}_{-0.05}}$, with a higher $N/M_\star=(5.4\pm 2.3)\times 10^{-3}$~M$_\odot^{-1}$. We have explored the implications of these DTDs for metal abundance evolution, both cosmic and stellar-Galactic. Our main findings follow.

\begin{enumerate}
 \item Using recent empirical estimates of the iron yields of the various SN types and of the observed relative mix of different CC SN types, we find that the average iron yield of a CC~SN is $0.074~{\rm M}_\odot$, and $0.7~{\rm M}_\odot$ for a SN Ia. Using these yields, and the SN rate evolution (based on the SFH and on the SN~Ia DTD) we have re-derived the cosmic iron abundance history. The cosmic mean iron abundance today is $8.1^{+1.3}_{-0.7}$\% of Solar, i.e., [Fe/H]=$-1.09^{+0.07}_{-0.04}$.
 
 \item $55\pm 6$\% of all iron has been made in SNe~Ia, and the rest in CC~SNe. This fraction is consistent with the fact that the lowest-metallicity stars, made of material that underwent little enrichment by SNe~Ia, have [Fe/$\alpha]\sim -0.4$, i.e., 40\% of Solar, meaning about 40\% of the Sun's iron is from CC~SNe.
 
 \item Under the valid approximation that almost all of the mass in $\alpha$ elements has been produced in CC~SNe, we have used the above, observation-based, SN~Ia and CC~SN contributions to the iron  accumulation history to derive the cosmic history of [$\alpha$/Fe]$(z)$. We have plotted the track of the mean cosmic volume element in the [$\alpha$/Fe] versus [Fe/H] plane. The track has a shallow descent up to a ``knee'' at $z\sim 2$. The knee is a result of the steep decline in the SFH (and the CC~SN rate) at $z<2$, combined with the shallower decline in the SN~Ia rate, because of the power-law tail of the DTD (and not because of the SNe Ia ``kicking in'' after their initial delay, as often claimed). This leads to a sharp increase in the relative SN~Ia contribution to the iron budget, and hence to the decrease in [$\alpha$/Fe].
 
 \item Using a very simplistic Galactic chemical evolution calculation, we have searched for effective Milky Way SFHs that, when combined with the above DTDs and iron yields, can reproduce the observed loci
 of stellar abundances in the [$\alpha$/Fe]--[Fe/H] plane. A SFH that peaks 10~Gyr ago, and declines by a factor 20 after 2~Gyr to a constant SFR level, reproduces well the observed low-$\alpha$ locus of MW stars usually associated with the thin disk. A single brief starburst $11.4$~Gyr ago (similar to the $z\sim 3$ star bursts that produced massive galaxy clusters), combined with the high-$N/M_\star$  ``galaxy-cluster'' DTD, gives an excellent match to the high-$\alpha$ locus of MW stars. This result points to distinct formation times and histories for these two MW components, as has been often noted
 before, but also to distinct SN~Ia production efficiencies, which we believe is a new result. In particular, star formation in the MW's halo and thick disk may have occurred at the same time and in a similar mode as star formation in massive galaxy clusters.
\end{enumerate}

Reconstructing the MW's SFH from stellar archaeology has a long tradition (e.g., \citealt{1979ApJ...229.1046T,1997nceg.book.....P,2001ASSL..253.....M}). Modern chemo-dynamical models have become increasingly sophisticated, as they attempt to implement numerous relevant physical processes, such as gas accretion and outflows, gas recycling by evolved stars, mixing of hot and cold gas, inter-zone 
radial mixing, the feedback of SNe and active galactic nuclei on the gas reservoir and on the star-formation rate, detailed metallicity-dependent element yields from SN models, and more. These efforts have been challenging, however. Many of the physical processes often included in such models cannot be guided by direct observations, nor can they actually be tested and constrained by stellar abundance data, as it is the highly degenerate product of many processes that, in the end, result in an effective star formation rate, one whose element yields are input into consecutive generations of stars. We have shown that the SN~Ia DTD and the iron yields of SNe are now consistently and sufficiently well determined that it is mostly just the effective MW SFH that needs to be invoked to match the observations in the [$\alpha$/Fe] versus [Fe/H] plane. Two simple effective SFH components, each with its particular empirical SN~Ia DTD normalization, succeed remarkably well in reproducing the two main observed concentrations of MW stars in this plane. This suggests that multi-component models and complex scenarios may not actually be necessary to explain stellar data. Nonetheless, the parameter space of the Galaxy's SFH still needs to be explored exhaustively, while fitting additional available stellar observables---ages, kinematics, space densities, and detailed abundance patterns---which we defer to future work.

\section*{Acknowledgments}

We thank Anna Ho, David Hogg, Piero Madau, Hans-Walter Rix, and Stan Woosley for discussions that stimulated and assisted this work.

D.M.'s work is supported by Grant 648/12 by the Israel Science Foundation (ISF) and by Grant 1829/12 of the I-CORE program of the PBC and the ISF. O.G. is supported by an NSF Astronomy and Astrophysics Postdoctoral Fellowship under award AST-1602595.

This research has made use of NASA's Astrophysics Data System and the NASA/IPAC Extragalactic Database (NED) which is operated by the Jet Propulsion Laboratory, California Institute of Technology, under contract with NASA.


\software{Matlab}


\end{document}